# THE SLOAN DIGITAL SKY SURVEY[†]

JAMES E. GUNN
*Dept. of Astrophysical Sciences, Princeton University*
*Princeton, NJ 08540, USA*

and

DAVID H. WEINBERG
*Institute for Advanced Study*
*Princeton, NJ 08540, USA*

ABSTRACT

We summarize the plans for and the current status of the Sloan Digital Sky Survey, a digital imaging and spectroscopic survey of $\pi$ steradians in the northern Galactic cap. The CCD photometric survey will produce images in five bands to limiting magnitudes of order 23. The spectroscopic survey will obtain redshifts of $10^6$ galaxies (a complete sample to a limiting magnitude $r' \sim 18$) and $10^5$ quasars ($g' \sim 19$). Repeated imaging of a 200 deg$^2$ strip in the southern Galactic cap will yield information about variable objects and a co-added photometric catalog roughly two magnitudes deeper than the northern survey. A dedicated 2.5-meter telescope, a large multi-CCD camera, and two fiber-fed double spectrographs are under construction and should be operational by fall of 1995. The main galaxy redshift sample will have a median redshift $\langle z \rangle \approx 0.1$.

astro-ph/9412080  20 Dec 94

## 1. Introduction

There are many fundamental, but as yet unanswered, questions about the structure of the Universe and the cosmogony of its constituents, from stars, through galaxies to quasars. Recent advances in mapping the local Universe provide a glimpse of the rich structure in the galaxy distribution. Yet we lack knowledge of even the lowest order statistical properties of the distribution of matter on large scales – the fluctuation spectrum of galaxies is poorly known on scales beyond 100 Mpc, a full decade below the comoving scale probed by COBE. Such a characterization is a necessary step to understanding the origins of large-scale structure. Our knowledge of the distribution and nature of the highest-redshift objects, quasars, is limited, as is our understanding of the intervening absorbing material. How and where galaxies form is in the realm of speculation, with continuing debate over the origins of the Hubble Sequence. Even the structure of our own Galaxy is controversial. In each of these cases, a major stumbling block is the lack of a uniform sample – of stars, galaxies or quasars – that has been reliably selected over a wide area of sky.

To produce the samples necessary to address these problems, a consortium of astronomers from what is now a large network of institutions is preparing to carry out a digital photometric and spectroscopic survey over a large fraction of the sky (10,000 deg$^2$) in the north Galactic cap, complete within precisely defined selection criteria, and a deeper imaging survey over 200 deg$^2$ in the southern Galactic hemisphere. In recognition of substantial funding from the Alfred P. Sloan Foundation, the project has been named the Sloan Digital Sky

---

Survey (SDSS). The participating institutions are the University of Chicago, Fermi National Accelerator Laboratory, Johns Hopkins University, the Institute for Advanced Study, the National Astronomical Observatory of Japan, Princeton University, the U.S. Naval Observatory, and the University of Washington.

The photometric map of the sky will measure accurate flux densities of objects almost simultaneously in five bands ($u'$, $g'$, $r'$, $i'$, and $z'$) with effective wavelengths of 3540 Å, 4760 Å, 6280 Å, 7690 Å, and 9250 Å, complete to limiting (5:1 signal-to-noise) point-source magnitudes ($AB_\nu$) of 22.3, 23.3, 23.1, 22.3, and 20.8, respectively. The planned sky coverage of about $\pi$ steradians will result in photometric measurements to the above detection limits for about $5 \times 10^7$ galaxies and a somewhat larger number of stars. The morphological and color information from the images will allow robust star-galaxy-quasar separation, yielding a photometric sample of about $10^6$ quasar candidates. Astrometric positions will be produced which we believe will be accurate to $\sim 50$ milliarcseconds for sources brighter than about 20.5. Medium resolution spectra will be obtained for the $10^6$ galaxies brighter than about $r' \sim 18$, approximately $10^5$ quasars brighter than $g' \sim 19$, and carefully selected samples of stars. The imaging survey in the south will go about 2.0 magnitudes deeper in all bands, and, in addition to being a bridge between the main northern survey and the much deeper, small-area surveys possible with very large telescopes, it will contain a wealth of information about faint variable sources, supernovae, and proper motions. The SDSS will make a substantial indirect contribution to all such very deep surveys by characterizing the nearby universe in a detailed and quantitative manner. Without this information, which does not exist in any satisfactorily accurate and complete form at present, one cannot compare the universe observed at great distances and ancient times to the universe today.

This talk briefly describes the survey instruments, the current status (November 1994) of the project, and the anticipated data products. We also discuss some of the expected properties of the main galaxy redshift sample. A more technically detailed description of the survey instruments and survey strategy appears in reference 1. Discussions of large-scale structure applications of the SDSS appear in the contributions of Vogeley and Weinberg elsewhere in this volume.

## 2. Instruments

The survey is defined in part by the capabilities of its instruments. From the early stages, it was clear that a project of this scope would need a dedicated, special-purpose telescope. We are constructing a 2.5-meter telescope with a modified Ritchey-Chrétien optical design that produces good images over a field of view 3° in diameter. The primary mirror is $f/2.2$ and the overall system is $f/5.0$. This design is well optimized for both a wide-area imaging survey and a multi-fiber spectroscopic survey of galaxies to $r' \sim 18^{\text{th}}$ magnitude.

The imaging camera contains 30 $2048^2$ CCDs, arranged in six columns. Each column occupies its own dewar and contains one chip in each of the five filters. The camera operates in drift-scan (TDI) mode: a star or galaxy moves down the column through the five filters, spending about 55 seconds in each. The pixel scale is 0.4". Columns are separated by about 83% of the active chip widths, so two interleaved scans cover a stripe of sky, with a 1' overlap region on either side of each column. The camera also includes 24 smaller CCDs arranged above and below the photometric columns, which are used for astrometric calibration of the imaging data and monitoring the focus. The photometric chips saturate at about $14^m$, but the astrometric chips have a much brighter saturation level because of neutral density filters and their smaller size, allowing the photometric data to be tied to the brighter stars in fundamental astrometric catalogs.

In addition to the 2.5-meter telescope, the survey will use an automated 0.7m telescope equipped with a CCD camera and the same filter set used in the main camera. This "monitor" telescope will observe a network of standard stars throughout every observing night to determine extinction correction coefficients and set zero points. These observations will be used to calibrate the photometry from the 2.5-meter imaging camera. The monitor telescope also has a set of narrow-band filters for use in spectrophotometric calibration.

The telescope will also be equipped with two double, fiber-fed spectrographs, each using two cameras, two gratings, and two $2048^2$ CCD detectors. The blue channel covers $\lambda\lambda\, 3900 - 6100$Å and the red channel $\lambda\lambda\, 5900 - 9100$Å. The spectral resolving power $\lambda/\Delta\lambda$ is about 1800 in the center of the wavelength range for both spectrographs. The optical fibers are 3" (180 microns) in diameter, and the two spectrographs together hold 640 fibers. The size of the fiber ferrules prevents us from obtaining simultaneous spectra of objects separated by less than 55"; when a close pair of galaxies falls in a region where plates overlap, we will get the second redshift on the second go-round. After exploring various complex robotic schemes for fiber positioning, we settled on the decidedly low-tech solution of plugging pre-drilled plates by hand. We plan to hire expert, dextrous fiber-pluggers rather than leave this crucial task to tired and impatient astronomers. We anticipate a spectroscopic exposure time of 30-45 minutes, with about 15 minutes of overhead per fiber plate. We will have ten fiber harnesses, so enough plates for a full night of beautiful weather can be plugged during the day.

It is worth noting that the combination of telescope and spectrograph is, in a sense, "over-engineered" for obtaining redshifts alone. We anticipate very high completeness in obtaining redshifts at the magnitude limit of the galaxy spectroscopic sample. For a large fraction of the galaxies, we will have spectra with the sort of signal-to-noise, resolution, and (especially) wavelength coverage that exist only for small samples today. These spectra will be valuable for stellar population and dynamical studies, especially when they are used in conjunction with the multi-color imaging data for the same galaxies. Velocity dispersions for elliptical galaxies can be used to study distance-indicator relations and large-scale flows. Similarly, the quasar spectra will provide detailed information about the quasars and about intervening material along the line of sight. At high redshifts, the SDSS spectra will not have sufficient resolving power to separate all the individual lines in the Lyman $\alpha$ forest, but properties of the Lyman $\alpha$ lines can still be investigated in some detail using the statistical techniques developed in the past few years for quasar absorption-line studies.[2,3,4]

The SDSS imaging survey will be the first large-area photometric survey to use CCD detectors. Why now? The venerable Palomar Sky Survey has been the backbone of many astronomical investigations for the past thirty years. Photographic plates lack the sensitivity, uniformity, linearity, and dynamic range of CCD detectors, resulting in biases and incompletenesses in photographic surveys. However, due to their large area, photographic plates have retained their competitive edge over CCDs. This situation is finally changing, because we now have the technology to produce working arrays of very large-area CCDs and the computing power to handle the extraordinary flood of digital data that such a detector system can produce.

The gain we achieve with our survey design may be quantified as follows: a measure of the information rate for an imaging survey is the *survey efficiency* $\epsilon$:

$$\epsilon = \Omega D^2 q,$$

where $\Omega$ is the solid angle of sky on the detectors, $D$ the diameter of the telescope, and $q$ the detector quantum efficiency. The time to complete a survey over a given area to a given limit and signal-to-noise ratio is clearly inversely proportional to $\epsilon$. The 48-inch Schmidt has an $\epsilon$ of about $0.30\,\mathrm{m}^2\,\mathrm{deg}^2$, allowing an (optimistic) photographic quantum efficiency of half a percent. The SDSS telescope with its CCD array will have an efficiency more than an order of magnitude larger, about 4.8 in these units, and it furthermore allows mapping the sky almost simultaneously in five bands.

The SDSS represents another new departure in that the photometric catalog for selection of the objects whose spectra will be measured will be done concurrently with the much more time-consuming spectroscopic survey. This strategy is designed to make optimum use of the observing time, as excellent photometric conditions (in terms of seeing, sky brightness and sky transparency) are present for only a minority of the time even at the best sites. The photometry will be done only in the best seeing conditions (median seeing at Apache Point, the telescope site, is about 0.8"). Spectroscopy will be carried out on less pristine nights. The hardware design allows for rapid changeover between photometric and spectroscopic modes, so that both can be done on the same night if the weather changes. This strategy allows the efficient and appropriate use of partial nights, including nights when the Moon is up for part of the night. The strip in the southern Galactic hemisphere will be observed when the north Galactic cap is inaccessible.

Execution of this strategy requires almost real-time production of a well-defined catalog of galaxies, quasars, and stars, to be achieved by reliable, fast software. The online data acquisition system at the mountain accepts data from the imaging camera, the spectrograph, and the monitor telescope, and writes two copies to tape. One set of tapes will be sent daily to Fermilab by courier for offline processing (exabytes on a plane offer a much higher baud rate than the Internet). Imaging data are reduced by three linked "pipelines": one for photometric calibration (using the monitor telescope data), one for astrometric calibration (using data from the astrometric chips), and the photometric pipeline itself, which uses these calibrations and the photometric CCD data to produce corrected pixel maps and catalogs of objects with measured, calibrated properties. Spectroscopic data are reduced by a separate pipeline, which extracts, calibrates, and classifies the spectra and measures quantities such as redshift, line strengths, and velocity dispersions. Additional support software is needed to keep track of the progress of the survey, to maintain the survey database, and to select spectroscopic targets from the imaging catalogs and generate instructions for drilling plates.

In imaging mode, the survey detector system generates raw data at 4.6 Mbytes per second, or 170 Gbytes on a 10 hour night. This data rate, and the associated requirement that the data reduction demand a minimal level of human intervention, place extraordinary demands on the survey software. The software is as ambitious and as essential to the project as the telescope, the camera, or the spectrographs, and it is receiving at least as much attention as these more obvious elements. The tasks of developing and testing the software are distributed among all eight of the participating institutions, with overall coordination and guidance from Fermilab.

## 3. Project Status

The hardware status of the project should have changed in significant ways by the time this volume appears. As of November, 1994, the telescope enclosure at Apache Point is built, and the monitor telescope is in place and will soon begin setting up the network of survey photometric standards. The 2.5-meter telescope is under construction and expected to be completed in March, 1995. We anticipate that the spectrographs, imaging camera, primary mirror, and secondary mirror will all be delivered to the mountain between March and August of 1995. We presently have in hand the four $2048^2$ CCDs for the spectrographs, all of the astrometric CCDs, and 15 of the 30 photometric CCDs for the imaging camera (not including 5 spares). We expect to have all of the photometric CCDs by January, 1996, at the latest; if necessary, we will operate the camera in the fall of 1995 with three of its six dewars.

If things proceed as expected, we should be able to commission instruments and start test observations in the fall of 1995, and we should have an operational system ready to produce data by January, 1996. The imaging survey does not have any "free parameters," so the imaging data that are taken under good conditions in 1996 will eventually become part of the survey data archive. For the galaxy and quasar spectroscopic surveys, however, we must finalize the magnitude and color criteria used to select targets from the imaging data, and we must "freeze" the photometric reduction software so that targets are selected in a uniform way over the lifetime of the survey. We anticipate that the spectroscopic observing during most of 1996 will be devoted to test programs, designed to check the robustness of our system and to help us decide on optimal selection criteria for galaxy and quasar spectroscopic targets. During this time, we will also test, debug, and improve the data reduction and survey control software. Toward the end of 1996, we expect to freeze the photometric software and the target selection criteria, re-reduce the raw imaging data that we have already taken, and begin the spectroscopic survey in earnest. Of course, our "test" programs should yield spectra of $50,000 - 100,000$ objects, and we expect to find some interesting science to do with them!

## 4. Data Products

The survey summarized in the introduction should take 5 years to complete, including the year (1996) devoted to imaging and test spectroscopic programs. At the end of the 5 years, there will be several different kinds of data products. The simplest and most manageable of these are the object catalogs, which contain lists of measured parameters for detected objects. The measured parameters will include positions, magnitudes, sizes, some simple morphological information, and radial profiles in a series of circular apertures. The parameter list for objects in the spectroscopic samples should be a pleasantly manageable 1-2 Gbytes, while the catalog of all objects in the photometric survey will be more unwieldy, 100-200 Gbytes. We are investigating ways to organize the catalog database so that it can respond to scientific queries in a sensible and efficient way.

At the opposite end from the object catalogs are the corrected pixel data (and the raw data themselves, which will also be saved), occupying something on the order of 10 Tbytes.

It will be possible but difficult to access the data at this level. We therefore plan to extract "atlas images" around all detected objects; we will attempt to make these large enough that nearly all science questions that deal with detected objects can be addressed by calling up their individual atlas images, without ever returning to the full pixel map. The size of the atlas image database depends on our as-yet-undecided compression scheme, but 200 Gbytes is a reasonable guess. We also plan to produce sky maps with various degrees of compression. Finally, the 1-d spectra ($\sim$ 50 Gbytes) and the flattened, 2-d spectroscopic frames ($\sim$ 70 Gbytes) will be saved.

The project is committed to releasing data from the survey's first two years of operation within two years of the time they are taken (i.e. four years from the start of the survey) and to producing a final public archive within two years of the completion of the survey. We anticipate that it will take us a significant fraction of the allotted time to ensure that the data are well calibrated and to develop practical ways of distributing such large quantities of information. The spectroscopic catalog data are more straightforward to deal with — certainly they are less voluminous — so we hope to distribute at least an early version of the first-year spectroscopic catalog on a somewhat shorter timescale. The final release will certainly involve some mix of distributed data and a central archive at Fermilab. The natural break will depend on developments in storage technology (CD-ROMs are the obvious distribution medium at present, but that could change in a few years) and network speeds (where, sadly, even the sign of future trends is not clear). In one attractive model, institutions could acquire their own copies of the object catalogs and atlas images on CD-ROM libraries, while lower level data would be archived at Fermilab.

## 5. Properties of the Galaxy Spectroscopic Sample

The main galaxy redshift survey will target the million brightest galaxies in the northern survey area. Over the past few months, we have begun to make a serious effort to define what we mean by "brightest." There are a number of desirable characteristics that we would like our spectroscopic target selection criteria to possess. They should be based on photometric parameters that we can measure accurately under the expected range of observing conditions, so that the final catalog will be uniform. They should be based on physically meaningful parameters. They should select a wide range of galaxy types. They should minimize the fraction of failed redshifts, or minimize the spectroscopic exposure time to reach a given completeness level. And, to the extent possible, they should be simple.

These desiderata do not always point in the same direction. For instance, isophotal magnitudes are simple to define and reasonably straightforward to measure, but it is not clear that they have a great deal of physical meaning. The isophotal radius in a galaxy of large total luminosity but low central surface brightness may enclose only a small fraction of the galaxy's light, or it may not be defined at all. Furthermore, as we move a given galaxy further away, the fraction of its light that contributes to an isophotal magnitude goes down steadily because redshift-dimming lowers the surface brightness, moving the isophotal radius in. Galactic extinction also changes the isophotal radius, so a proper extinction correction requires knowledge of the full galaxy profile. Selection based on 3" aperture magnitudes might yield the highest completeness for fiber redshifts, but such a magnitude seems even more dubious than an isophotal magnitude from a physical point of view, and it is sensitive to seeing. We would prefer to select on something like a total magnitude, but all methods of measuring total magnitudes require some kind of model-based extrapolation to account for light that is lost in the galaxy's noisy outer regions.

We are presently planning to use galaxy magnitudes based on a modified form of the Petrosian[5] radius, which is a radius at which the ratio of the local surface brightness to the averaged interior surface brightness falls to a specified value. Specifically, we define the Petrosian radius $R_P$ implicitly by the relation

$$\frac{\bar{I}_{\text{ann}}}{\bar{I}_{\text{int}}} \equiv \frac{\int_{0.8 \times R_P}^{1.25 \times R_P} I(r) 2\pi r\, dr / [\pi(1.25^2 - 0.8^2)R_P^2]}{\int_0^{R_P} I(r) 2\pi r\, dr / [\pi R_P^2]} = f_1,$$

where $I(r)$ is the object's azimuthally averaged surface brightness in the band under consideration. Note that the "local" surface brightness is in fact averaged over a fairly wide annulus, in order to suppress effects of noise and small-scale structure within the galaxy. We define the Petrosian magnitude by the light inside a fixed number of Petrosian radii:

$$m_P \equiv -2.5 \log \left\{ \frac{\int_0^{f_2 \times R_P} I(r) 2\pi r\, dr}{3630\,\text{Jy}} \right\}\,.$$

The 3630 Jy normalization puts us on the AB magnitude system. The free parameters are $f_1$, the surface-brightness ratio used to define the Petrosian radius, and $f_2$, the number of Petrosian radii over which the light is integrated; we have tentatively adopted the values $f_1 = 1/8$ and $f_2 = 2$. With these parameters, the Petrosian magnitude incorporates 90% of the light from an object with an $R^{1/4}$-law profile and essentially all of the light from an object with an exponential profile; these numbers are insensitive to the axis-ratio of the galaxy, even though we define the Petrosian quantities in circular apertures.

The Petrosian magnitude has most of the advantages of a "total" magnitude, and it does not require model-dependent extrapolations for missed light. The one disadvantage of the Petrosian magnitude for our purposes is that it does not directly tell us an operationally important quantity: the amount of light that will go down a 3-arcsec fiber. To avoid wasting time on galaxies that will not yield a fiber redshift, we plan to supplement the Petrosian magnitude limit with a surface-brightness threshold. We will probably base this threshold on the 1/2-light surface brightness, i.e. the mean surface brightness inside a circular aperture that encloses half of the light contributing to the Petrosian magnitude. We plan to do all of our galaxy selection in $r'$ band; $g'$ magnitudes would have similar signal-to-noise but would be more sensitive to Galactic extinction, while $i'$ magnitudes would be more subject to variations in sky brightness.

We have been making use of simulations in a number of different areas of the project, including our investigations of target selection. Our galaxy catalogs are drawn from a large N-body simulation of a low-density CDM model ($\Omega = 0.4$, $\Lambda = 0.6$, $h = 0.6$), performed for this purpose by Changbom Park and Richard Gott. The simulation volume is a $600h^{-1}\,\text{Mpc}$ periodic cube containing 54 million particles, of which a biased subset of 8 million represent "galaxies." We assign galaxies absolute B-magnitudes drawn from a Schechter luminosity function, morphological types selected in accord with the observed morphology-density relation, diameters, bulge-to-disk ratios, and axis ratios based on observed distributions (often quite uncertain), and colors in the SDSS bandpasses based on published spectrophotometry of galaxies of different Hubble types. Figure 1 shows a 6° slice through our mock redshift catalog. We adopt a Petrosian magnitude limit $r' < 18.03$ and an $r'$ surface-brightness threshold of 22 mag/arcsec$^2$, which yields 973,204 galaxies over the $\pi$-steradian survey area. The surface-brightness threshold eliminates 20% of the simulated galaxies that are brighter than the Petrosian magnitude limit, cutting off the faint tail of the 3" aperture magnitude distribution.

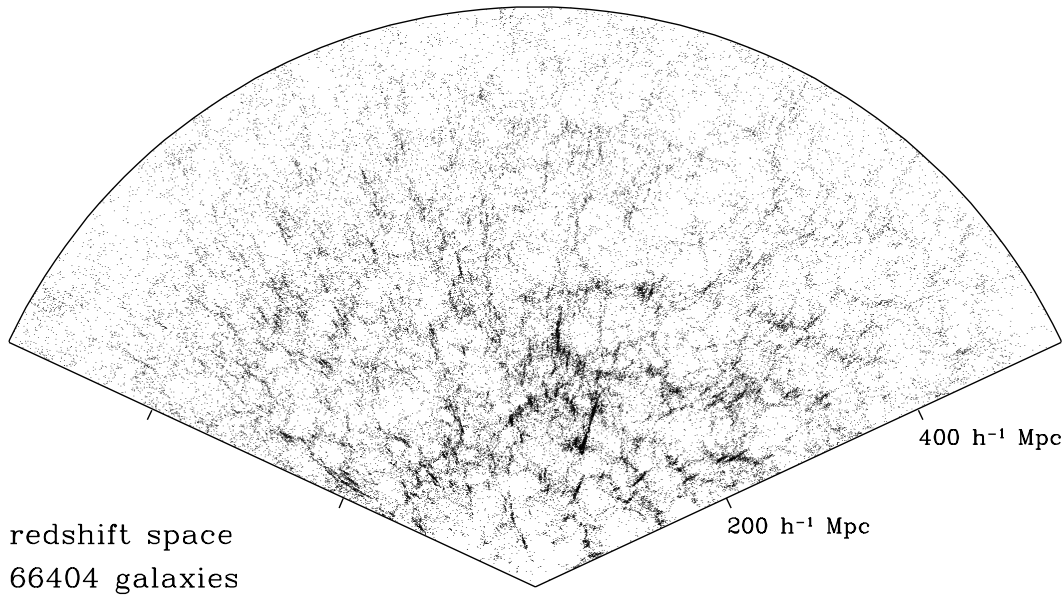

Figure 1: A simulated slice, 6° by 130°, through the SDSS redshift survey of the north Galactic cap. Galaxies are plotted at the distance that would be inferred from their redshift, so cluster velocity dispersions create "fingers of God" that point towards the observer. The slice contains 66404 galaxies, 6.6% of the number expected over the full area of the northern survey. This mock catalog is drawn from a large N-body simulation of a low-density ($\Omega = 0.4$, $\Lambda = 0.6$) CDM model, performed by Changbom Park and Richard Gott.

Figures 2 and 3 show the redshift distribution and associated selection function derived from this mock redshift catalog. These plots represent our best current guess about the distribution of redshifts and types in the spectroscopic sample. However, the simulations are based on assumptions about the galaxy population; some of these assumptions have solid empirical backing, but many of them will not be adequately tested until the SDSS data themselves become available. This is one reason that the "test year" mentioned in §3 is so essential to the project. The test year will play an even more important role in devising quasar selection criteria, since our knowledge of the quasar and stellar loci in the SDSS color system is quite incomplete.

In addition to the northern spectroscopic samples, we will have redshifts of galaxies and quasars in the southern strip (and in two other strips in the southern Galactic hemisphere). The southern galaxy redshift sample will not go much deeper than the northern sample — a 2.5-meter telescope is not the right instrument for getting faint galaxy redshifts — but we will push somewhat beyond the selection boundaries of the northern sample so that we have a better idea of what galaxies we are missing. The southern quasar sample will go 1-2 magnitudes deeper than the northern sample, providing better information about small-scale quasar clustering and the low end of the luminosity function. A significant chunk of the southern spectroscopic time will go to a sample of objects brighter than $g'_{3"} \sim 19$, with no other color or morphological selection. Most of these objects will be stars, and the accurate spectral types, abundance measurements, and radial velocities for such a large sample will make important contributions to our understanding of the stellar populations and dynamics of the Galaxy. Some of the objects will be compact galaxies and quasars with near-stellar colors. Above all, we hope that taking spectra in a very large haystack will turn up a few extremely interesting needles.

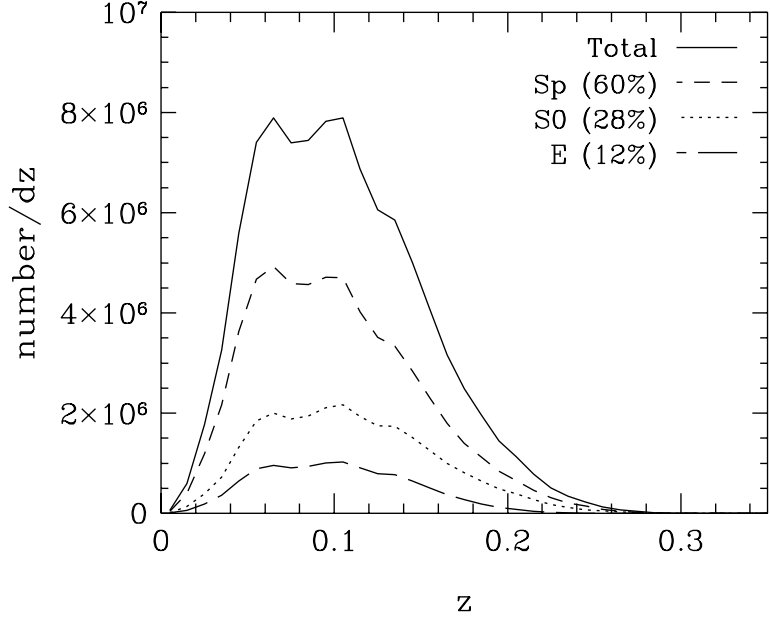

Figure 2: Redshift distributions of the simulated sample illustrated in Figure 1, in number of galaxies per unit redshift over the $\pi$-steradian survey area.

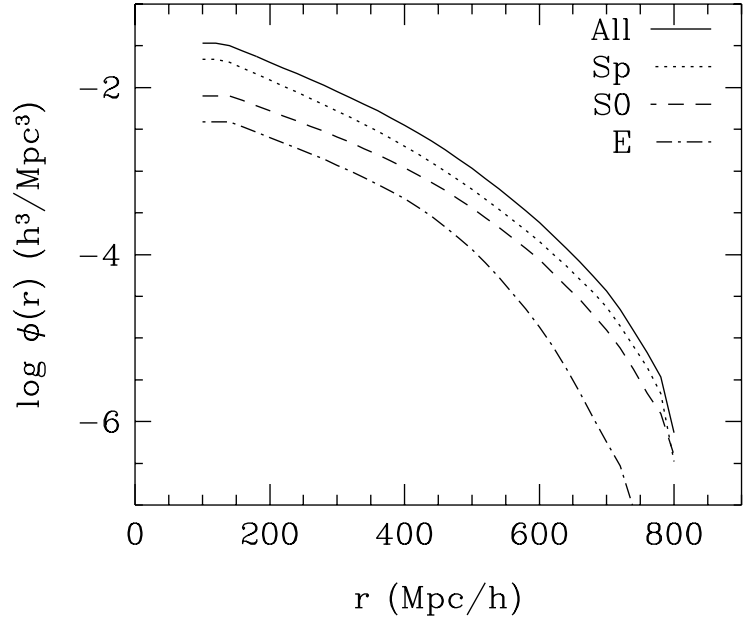

Figure 3: Selection functions derived from the simulated redshift sample; $\phi(r)$ is the mean number of galaxies per $(h^{-1}\mathrm{Mpc})^3$ expected at distance $r$. The absence of very faint galaxies in the simulated sample causes $\phi(r)$ to underestimate the value expected for the real survey inside about $200\,h^{-1}\mathrm{Mpc}$. The drop in the elliptical selection function beyond $500\,h^{-1}\mathrm{Mpc}$ may arise from spatial structure in the mock catalog.

## 6. Concluding Remarks

One of the major motivations for the SDSS is, of course, to study the large-scale structure of the universe and to test theories for its origin. Several factors will make the SDSS a uniquely powerful database for studying large-scale structure. One is the sheer number of galaxies in the redshift sample, nearly a two-order-of-magnitude increase over the largest surveys that exist today. Another is the quality and uniformity of the photometric data that will be used to select spectroscopic targets and measure Galactic extinction; these will give superb control over systematic effects that might otherwise limit the accuracy of clustering analyses on large scales. The photometric data and high-resolution galaxy spectra also allow one to classify the galaxies of the redshift sample in a variety of ways, making it possible to study the relative clustering of different galaxy types. Finally, while the 1,000,000 galaxy redshift sample is perhaps the most visible element of the SDSS, at least from the point of view of large-scale structure, the deeper, multi-color, photometric survey and the spectroscopic survey of quasars and their absorption systems will also make major contributions to our understanding of galaxy and structure formation.

The SDSS data will support a broad range of investigations outside of large-scale structure. The "survey science" section of our recent NSF proposal ran to 99 pages, and that included only the ideas that members of the collaboration had the imagination to think of and the energy to write down. If the survey performs to our expectations and hopes, many of its most exciting results will be orthogonal to the things we have already thought about, and the data themselves will be useful long after the papers that emerge in the next few years have made their way to the dungeon of compact shelving.

From the perspective of two deeply immersed optimists, the project seems to be going remarkably well. We have had our share of frustrations and delays, but none have been catastrophic, or beyond what one might reasonably expect. All of us are learning both the rigors and the joys of working in a large, diverse collaboration. A month after the *Wide-Field Spectroscopy* workshop, the Sloan project held its own, 10-day, internal workshop at Yerkes Observatory, which was attended by more than 60 scientists from the survey institutions. Our discussions at this workshop made clear the enormity of the tasks that lie in front of us, but they also showed us that we have come a long way already, and that we have a large and talented team to take us ahead. We are *very* eager to see the first images from our camera and the first spectra from our spectrograph. We hope that readers of this volume will find wonderful science to do with the astronomical quantities of data that we are preparing to push your way.